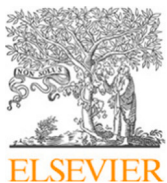
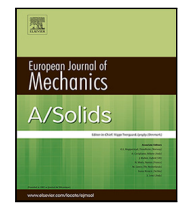

Full length article

# Phenomenological modeling of the stress-free two-way shape-memory effect in semi-crystalline networks: Formulation, numerical simulation, and experimental validation

Matteo Arricca [a], Nicoletta Inverardi [b], Stefano Pandini [b], Maurizio Toselli [c], Massimo Messori [d], Ferdinando Auricchio [a], Giulia Scalet [a],*

[a] *Department of Civil Engineering and Architecture, University of Pavia, via Ferrata 3, Pavia 27100, Italy*
[b] *Department of Mechanical and Industrial Engineering, University of Brescia, via Branze 38, Brescia 25133, Italy*
[c] *Department of Industrial Chemistry "Toso Montanari", University of Bologna, Viale Risorgimento 4, Bologna 40136, Italy*
[d] *Department of Applied Science and Technology, Politecnico di Torino, Corso Duca degli Abruzzi 24, Torino 10129, Italy*

ABSTRACT

Polymers exhibiting the stress-free two-way shape-memory effect (SME) represent an appealing solution to achieve self-standing reversible actuation that is a fundamental feature required by numerous applications. The present paper proposes a one-dimensional continuum phenomenological framework to model single-component semi-crystalline polymer networks exhibiting both the one-way SME and the two-way SME under stress and stress-free conditions. A comprehensive experimental campaign is first performed on semi-crystalline networks based on poly($\varepsilon$-caprolactone) (PCL) to characterize the mechanical and thermal properties as well as the one-way and two-way shape memory behavior of the material under different thermo-mechanical conditions. The results guide the formulation of the model, elucidating the selection of the control and phase variables and motivating the choice of their evolution laws. Model capabilities are then demonstrated against experimental data. All the phenomena that influence the stress-free two-way SME, including the actuation temperature, heating/cooling rates, applied stress/strain, and the amount of skeleton and actuation phase, are analyzed and discussed, giving new important insight for application development.

## 1. Introduction

Shape-memory polymers (SMPs) are stimuli-responsive materials capable of recovering a *permanent* shape, provided via material processing, from a previously imposed *temporary* shape. Such an outcome is termed shape-memory effect (SME) and occurs through a specific shape-memory cycle, wherein SMPs experience an external stimulus such as heat, light, or a magnetic field. SMPs display advantageous properties, *e.g.*, low cost, tunable behavior, low weight, high deformability, and biocompatibility, and therefore attract significant interest in various application fields, ranging from biomedical and pharmaceutical (Xia et al., 2020), to aerospace (Scalet, 2020). Recently, the combination of SMPs with additive manufacturing techniques, often referred to as 4D printing, has also opened great perspectives in the fabrication of smart components with complex, personalized shapes and high precision.

Among others, the thermally-triggered one-way SME is the most studied effect. It is ensured by a proper combination of the polymer macromolecular architecture and the applied thermo-mechanical history (known as shape-memory cycle) and strictly depends on the transition temperature (*e.g.*, melting or glassy) as a function of the polymer type. It is nonetheless worth to point out that the one-way SME is a non-reversible feature, in view of the necessity of resorting to an external mechanical intervention to set again the (programmed) temporary shape from the (recovered) permanent one. As a consequence, such an effect is not suitable for applications where a reversible behavior is demanded, as in the fields of soft actuators and robotics.

Differently, the two-way SME ensures reversibility between two configurations upon the application of a cooling–heating cycle. Early works (Basak and Bandyopadhyay, 2022; Feng and Li, 2022) focused on the two-way SME under an external applied stress in liquid crystalline polymers and in semi-crystalline (cross-linked polymer) networks. Here, the presence of both a crystallizable phase and chemical cross-links is necessary to provide the two-way SME, that is generally obtained by means of cooling–heating cycles from above the melting temperature to below the crystallization temperature under an external

---

applied stress. Recently, it has been shown that the two-way SME does not necessarily require the presence of an external applied stress, leading to the possibility of a self-standing (stress-free) reversible actuation. Indeed, as detailed below, in the case of polymers with specific macromolecular architectures and thermo-mechanical histories, this two-way SME arises from a generated internal stress and takes the name of *reversible bidirectional* or stress-free two-way SME.

The achievement of the reversible bidirectional SME is allowed by the presence of two phases that work synergistically: a skeleton phase, which provides the internal stress necessary, and an actuation phase, which is responsible for the actuation by undergoing a reversible shape change upon crystallization-melting cycles under the generated internal stress.

The increasing number of applications requiring self-standing reversible actuation motivates a comprehensive understanding of the stress-free two-way SME and of its mechanisms and phenomena behind, as well as the development of accurate models and numerical tools for design purposes.

Among the various strategies that may be conceived to induce reversible two-way SME, as well depicted in Wang et al. (2019), three categories based on the employed polymer macromolecular architecture may be identified, hence, (*i*) homo-polymers, single-component semi-crystalline networks presenting a unique type of crystalline phase (Turner et al., 2014; Tippets et al., 2015; Dolynchuk et al., 2017; Xu et al., 2019; Yuan et al., 2020; Wong et al., 2023; Behl et al., 2013; Zhou et al., 2014; Wang et al., 2017, 2020; Xu et al., 2022; Wang et al., 2022; Hao et al., 2022; Huang et al., 2023; Ren and Feng, 2023), (*ii*) multiphase semi-crystalline networks (Saatchi et al., 2015; Fan et al., 2018; Yan et al., 2018; Inverardi et al., 2022; Li et al., 2022; Jiang et al., 2022; Xiang et al., 2022; Uto et al., 2023), and (*iii*) networks obtained by a secondary post-curing treatment performed on a stretched structure (Meng et al., 2015).

The present work moves from this motivation and focuses on SMPs of category (*i*). The simple structure of homo-polymer networks provides an easy achievement of the SME, avoiding the complex multiphase structures which require non-trivial polymer synthesis and characterizes polymers of the aforementioned categories (*ii*) and (*iii*). In (*i*), the SME derives from an interplay, upon partial melting at a certain temperature (named actuation temperature), between the unmelted and melted chains that act as skeleton and actuation phase, respectively. It is worth to point out that the actuation temperature has to be carefully controlled in order to avoid either the complete melting of the skeleton, if too high, or an unclear production of the two-way effect, for too low temperatures.

Despite the synthesis of homo-polymers has been object of exhaustive studies, comprehensive thermo-mechanical characterization and modeling of the stress-free two-way SME still require deep investigations. Indeed, except for one contribution (Yan et al., 2020), most of the models were proposed for describing the two-way SME under an applied non-zero stress in a one-dimensional (Westbrook et al., 2010; Dolynchuk et al., 2014, 2015; Scalet et al., 2018) and three-dimensional (Zeng et al., 2021; Prasad et al., 2021; Zeng et al., 2023) framework. To the best of authors' knowledge, all the phenomena that may influence the stress-free two-way SME, including the actuation temperature, heating/cooling rates, applied level of stress/strain, and the amount of skeleton and actuation phase, have not yet been investigated. Understanding and predicting these phenomena are of fundamental importance to propose and design new solutions for innovative applications as soft robotics.

The aim of the present work is to provide a one-dimensional phenomenological continuum framework to characterize the SMP behavior of networks belonging to the aforementioned category (*i*). Particularly, a phase transition approach, first proposed by Liu et al. (2006) for amorphous polymers, is adopted thanks to its simplicity and generality. In fact, the approach has been applied to a wide variety of SMPs, also coupled with physically-based phase evolution laws (Guo et al., 2016; Yang and Li, 2016). Accordingly, the microstructure of the SMP network under investigation is represented, at the macroscopic scale, by phase parameters describing the amorphous soft phase, the crystalline hard phase, and the inelastic contributions associated with the shape-memory deformation. Evolution laws and constitutive equations are subsequently derived on the basis of physical evidences. A comprehensive experimental campaign on semi-crystalline networks based on poly($\varepsilon$-caprolactone) (PCL) is performed for model formulation and calibration, encompassing new insights on SMP material behavior, and with particular focus on the crystallization evolution under full and partial melting. All the phenomena that may influence the stress-free two-way SME are investigated and discussed.

The proposed model is characterized by the advantage of being straightforward and simple to implement, and it is based on physically interpreted and of easily measurement parameters provided by the well founded and widely adopted testing protocols for SMPs. Model capabilities are assessed through several numerical tests, reproducing both the one-way and the two-way SME under different thermo-mechanical histories under strain- and stress- control.

The manuscript is organized as follows. Section 2 presents the results of the conducted experimental campaign. The correlations between experimental evidences, their physical interpretation, and the model formulation are discussed in Section 3. Section 4 presents the formulation of the continuum model and its numerical implementation, whereas Section 5 shows model calibration and the comparison between numerical and experimental results to validate the model. Section 6 completes the paper discussing conclusions and future perspectives.

## 2. Material and experiments

This section presents the results of the conducted experimental campaign, starting from material synthesis and thermo-mechanical properties, and providing a characterization of both the one-way and two-way shape-memory behaviors.

### 2.1. Material synthesis

Semi-crystalline PCL network was prepared to start from the following products: poly(caprolactone)-diol (Mn: ~10 kDa); 2-isocyanatoethyl methacrylate (2-IEM, 98%); Tin(II) 2-ethylhexanoate; tetrahydrofuran (THF); 2-hydroxy-2-methyl-1-phenyl propanone (Additol HDMAP). PCL-diol, 2-IEM, and Tin(II) 2-ethylhexanoate were purchased from Sigma-Aldrich and used as received, without any further purification, while Additol HDMAP was purchased from Cytec.

The synthesis involved first the metacrylation of the hydroxyl end groups of PCL with 2-IEM, added with a 20% stoichiometric excess with respect to hydroxyl groups of the diol macromonomers: after drying the polymer at 60 °C under dynamic vacuum in the presence of molecular sieves, methacrylate end-capping of hydroxy-terminated PCL was carried out by reacting them with 2-IEM in bulk at 100 °C for about 3-4 h in presence of organotin catalyst, under nitrogen atmosphere and mechanical stirring. Unreacted 2-IEM was removed by dynamic vacuum at the end of the reaction. The reaction was monitored by IR until the ratio of peak strength at 2930 cm$^{-1}$ to 2275 cm$^{-1}$ became constant and to ensure the absence of 2-IEM, which is confirmed by the lack of the signal corresponding to isocyanate group at 2275 cm$^{-1}$.

Cross-linked PCL was prepared by carefully mixing the dimethacrylated PCL at 80 °C with the radical photoinitiator (2-hydroxy-2-methylpropiophenone; 0.1% wt). The well-mixed melts were poured between two glass plates (25 mm × 75 mm) and Teflon spacers, and photocross-linked on a heated plate under 365 nm UV irradiation (Hamamatsu LC8 spot light source; intensity 6 mW cm$^{-2}$) held at 12 cm above the sample for 5 min on both sides. Free films with thickness of about 300-600 μm were obtained by peeling them from the glass plate and completing the curing process by irradiating further 5 min on each side.

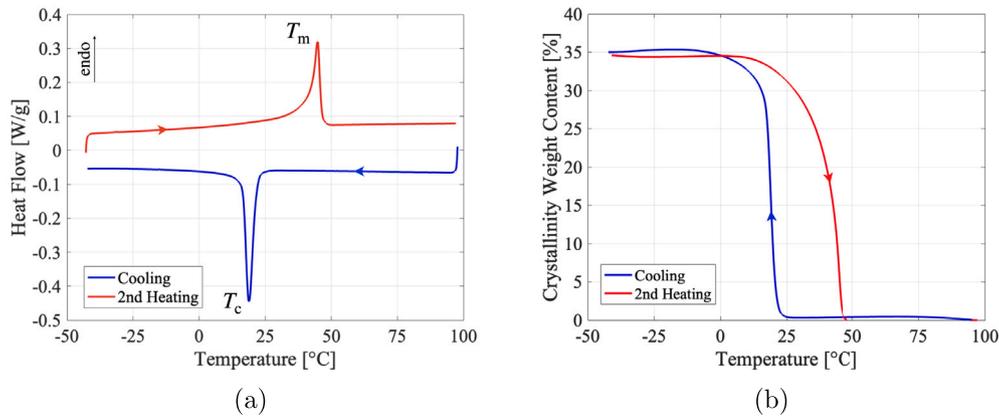

**Fig. 1.** (a) DSC traces (cooling and second heating) and (b) evaluation of the crystallinity weight content change upon complete crystallization and melting.

## 2.2. Thermal and mechanical properties

Differential scanning calorimetry (DSC) was carried out by means of a DSC Q10 (TA Instruments), according to the following program: (*i*) heating from −40 °C to 100 °C at 2 °C/min; (*ii*) cooling to −40 °C at 2 °C/min; (*iii*) heating to 100 °C at 2 °C/min. The crystallinity weight evolution upon cooling and upon the second heating scan was measured. Particularly, the weight crystal content, $\xi_C^w$, was calculated according to the relations

$$\xi_C^{w,\text{cool}}(T) = \frac{\Delta H^{\text{cool}}(T)}{\Delta H_{100}}, \qquad (1)$$

for the trace measured upon cooling, and

$$\xi_C^{w,\text{heat}}(T) = \frac{\Delta H^{\text{heat}}(T)}{\Delta H_{100}}, \qquad (2)$$

for the trace measured upon heating. Here, $\Delta H^{\text{cool}}(T)$ represents the exothermal enthalpy evaluated as a function of the decreasing temperature, and was evaluated upon crystallization as the area between the exothermal trace and a properly defined baseline, while $\Delta H_{100}$ stands for the theoretical enthalpy of a 100% crystalline PCL (*i.e.*, equal to 135 J g$^{-1}$ (Crescenzi et al., 1972)). $\Delta H^{\text{heat}}(T)$ represents the enthalpy evaluated upon melting as a function of the increasing temperature, and was evaluated as the area between the endothermal trace and a properly defined baseline.

DSC traces of the cooling and second heating run are reported in Fig. 1(a), whereas the dynamic crystal formation and melting upon cooling and heating, as deduced from the curves in Fig. 1(a), are represented in Fig. 1(b). The result allows us to evaluate the characteristic temperatures of crystallization and melting (respectively, $T_c = 19$ °C and $T_m = 45$ °C), as well as the maximum crystallinity content, which is about 35%. The crystallinity evolution along the cooling–heating cycle induces a crystallization, occurring as a sigmoidal increase as a function of temperature, onsetting at about 25 °C, crystallizing up to 80% in few degrees, but needing to be cooled down to about −10 °C to achieve the crystallization of the remaining 20%. These crystals are stable up to 10 °C and then a broad pre-melting process onsets up to 30 °C, becoming significant at 45 °C and finally being completed right before 50 °C.

Interestingly, DSC was also employed to investigate the crystallinity evolution in cyclic heating–cooling tests, promoting partial melting and subsequent crystallization cycles in a similar manner as in the occurrence of reversible two-way shape memory cycles. A material specimen was subjected to a cyclic thermal history according to the following steps: (1) after a first heating at 100 °C to promote complete melting, full crystallization occurs by cooling at −40 °C at 2 °C/min; (2–3) partial melting by heating at 45 °C at 5 °C/min, followed by crystallization at −40 °C at 5 °C/min; (4–5) partial melting by heating at 47.5 °C at 5 °C/min, followed by crystallization at −40 °C at 5 °C/min; (6–7) partial melting by heating at 50 °C at 5 °C/min, followed by crystallization at −40 °C at 5 °C/min; (8) full melting by heating at 100 °C at 5 °C/min.

The DSC traces of the various steps are represented in Fig. 2(a), while the corresponding dynamic crystal formation and melting are reported in Fig. 2(b). The results of the pairs (2–3), (4–5), and (6–7) describe the crystallinity evolution in cycles of partial melting and subsequent crystallization of the melted crystals, whereas the traces (1) and (8) may be considered as referred to a full melting-crystallization cycle. The apparent gap between cycles (4) and (5), and cycles (6) and (7), is due to difficulties in providing proper heat flow data when moving from the heating to the cooling phase, due to thermal inertia. In the partial melting-crystallization cycles the DSC traces appear as portions of the melting peak, followed by a progressively larger crystallization process as the maximum cycle temperature increases. Furthermore, while all the melting processes occur at the same temperature, the crystallization processes take place at higher temperature in presence of larger crystalline fractions, as a consequence of a more effective nucleation due to the presence of the remaining crystalline phase.

The mechanical behavior of the material is strongly influenced by the presence of crystals, moving from a glassy rigid behavior below $T_c$, to a rubbery behavior above $T_m$. Tests were carried out by means of dynamic-mechanical analyzer (DMA Q800 - TA Instruments) in order to describe the evolution of the mechanical properties with temperature. Particularly, tests were carried out by applying a cyclic cooling and heating cycle under tensile conditions on rectangular strips (gauge length 10 mm; width: about 5 mm), at a frequency of 1 Hz and at a rate of 2 °C/min, first by heating at 65 °C to promote full melting. Results are represented in Fig. 3 in terms of storage modulus as a function of temperature. The storage modulus shows a reversible change between a glassy value of about 300 MPa and a rubbery one of about 5 MPa. The change between the values occurs along a sigmoidal curve, whose inflection point is right above $T_m$ in the heating scan and along the cooling run at about $T_c$. Outside the sigmoidal part of the curve, a slightly increasing dependence on temperature may be observed at high temperature, consistently with rubber elasticity, while a decreasing dependence is found at low temperature, and this is ascribed to a progressive reduction of the material density, which leads to an decrease in material stiffness.

The mechanical behavior was evaluated above $T_m$, in tensile tests carried out on the DMA Q800 on rectangular strips (gauge length 10 mm; width: about 5 mm). A tensile ramp was applied at 80 °C and at a testing rate of 1 N min$^{-1}$, with the aim to evaluate material stress vs strain behavior. The results are reported in Fig. 4, showing a typical non-linear rubbery response and highly compliant behavior. The material stiffness at this temperature was evaluated in the linear slope of the curve (about 4.55 ± 0.16 MPa), and the material never above $T_\mathrm{m}$; ($ii$) applying a given strain; ($iii$) cooling below $T_\mathrm{c}$ under fixed strain conditions, followed by unloading; ($iv$) heating above $T_\mathrm{m}$ under stress-free conditions. This cycle was applied to rectangular strips (length between grips: 10 mm, width: about 5 mm), under tensile loading, by means of DMA Q800.

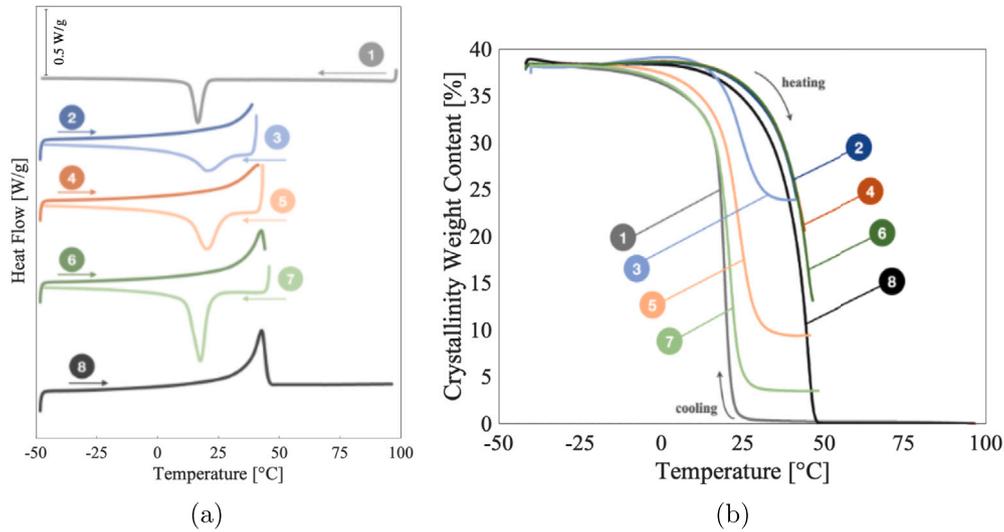

**Fig. 2.** Crystallinity weight content, evaluated in DSC tests, along the following cyclic heating–cooling program: (1) after a first heating at 100 °C, cooling at −40 °C; (2) heating at 45 °C; (3) cooling at −40 °C; (4) heating at 47.5 °C; (5) cooling at −40 °C; (6) heating at 50 °C; (7) cooling at −40 °C; (8) heating at 100 °C.

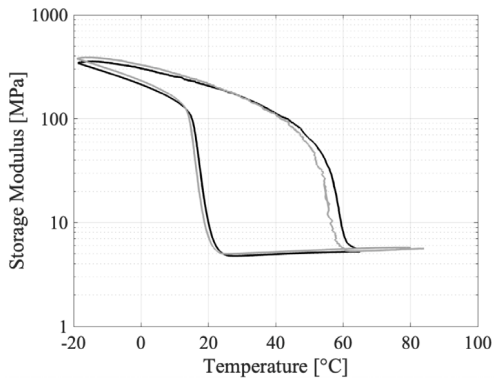

**Fig. 3.** Storage modulus as a function of temperature.

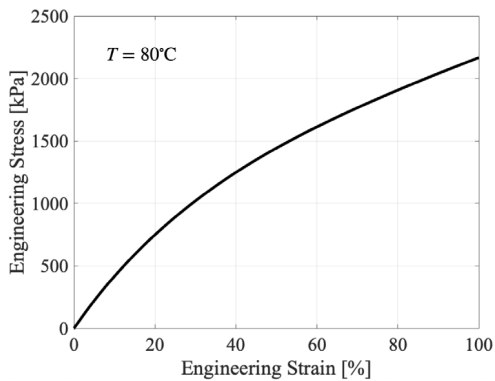

**Fig. 4.** Stress vs strain curve for the material in the rubbery plateau.

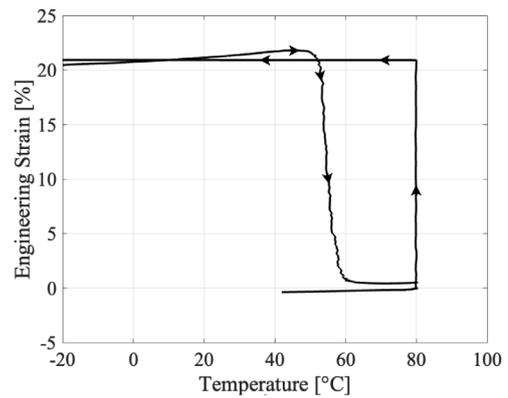

**Fig. 5.** Strain vs temperature curve demonstrating the one-way SME. Arrows indicate the direction in which steps ($i$)-($iv$) are applied.

The obtained strain-temperature response is represented in Fig. 5. In general, the material displayed a very high strain fixity, *i.e.*, the percent of the applied strain that is maintained after unloading in step ($iii$), and a very high strain recovery in step ($iv$). During step ($iv$), the slight strain increase below $T_\mathrm{m}$ and right before the narrow strain recovery across the melting region, is associated to thermal expansion.

### 2.4. Two-way shape-memory behavior

The two-way shape memory behavior was investigated on rectangular strips (length between grips: 10 mm, width: about 5 mm) by means of the DMA Q800 in tensile mode under three different testing protocols in order to explore the influence of various thermo-mechanical parameters.

#### 2.4.1. Non-zero stress conditions

In the first protocol, a tensile deformation equal to 20% was applied to the specimens at 80 °C and the shape-memory behavior was studied displayed failure within the tensile ramp at least up to a strain equal to 80%.

### 2.3. One-way shape-memory behavior

The one-way shape-memory behavior of the material was characterized following the standard thermo-mechanical cycle employed for semi-crystalline materials. It consists of the following steps: ($i$) heating under cooling–heating cycles under a constant stress (corresponding to the 20% of strain) between −20 °C and 80 °C at 2 °C min$^{-1}$. Strain-temperature curves are reported in Fig. 6(a). As it can be observed, the elongation–contraction in response to the cooling–heating cycle carried out under the application of a non-zero tensile fixed stress is confirmed: during cooling, the material elongates continuously, with a steeper increase across the $T_c$ region (known as cooling-induced elongation, CIE), up to a plateau; during heating, the material system undergoes a contraction across the $T_m$ region (known as melting-induced contraction, MIC), which, in dependence of the material and testing conditions, leads to a complete (or almost complete) recovery of the strain. It should be noted that, according to previous investigations (Scalet et al., 2018), some further effects may be reported: (*i*) a higher value of the applied stress leads to a larger extent of the elongation–contraction cycle; (*ii*) a threshold value for the stress is required to have a net elongation, since for lower stresses the elongation, although present, is of similar entity of the concurrent thermal contraction; (*iii*) the inflection point of the sigmoidal increasing trend of the CIE moves to higher temperatures as the applied stress increases. This latter effect can be attributed to the easier crystallization of stretched and aligned PCL chains, and regards only the CIE. The thermal loading conditions also play a relevant role: higher heating rates lead to a steeper MIC process and to a shift to higher temperatures of the overall MIC effect; similarly evidenced for the heating step, higher cooling rates lead to a steeper strain increase and to a shift to lower temperatures of the CIE.

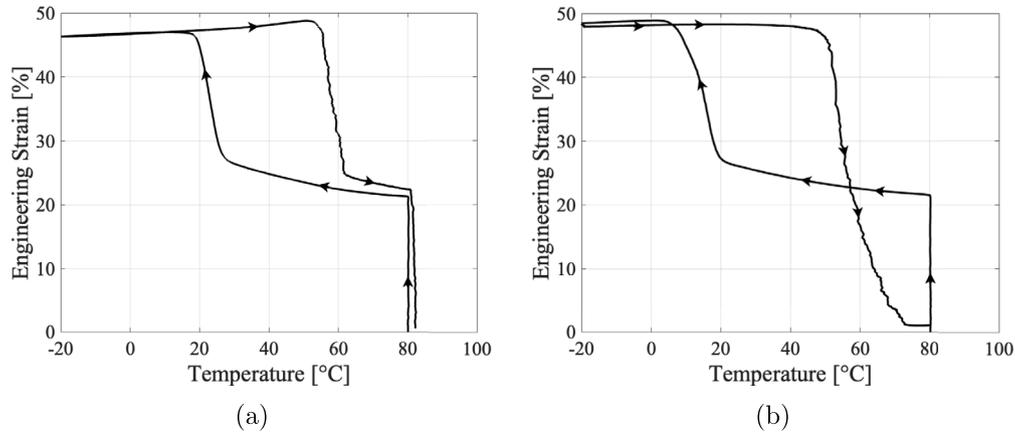

**Fig. 6.** Strain vs temperature curve demonstrating the two-way SME under (a) cooling–heating cycles under constant applied stress and (b) stress-free recovery. Arrows indicate the direction in which steps are applied.

Fig. 6(b) displays the capability of the material to recover its original shape after the CIE, if heated under stress-free conditions.

### 2.4.2. Zero stress conditions

The stress-free reversible two-way SME was investigated with the aim of understanding the effect of the applied pre-strain and of the actuation temperature of the cycle.

As discussed in Section 1, the effect relies on a specific thermo-mechanical procedure which involves the following steps: (*i*) deforming the polymer above $T_m$ and cooling it below $T_c$ under deformed condition (either under fixed stress or fixed strain conditions) until complete crystallization; (*ii*) after having removed the applied stress, heating the polymer up to a so-called actuation temperature, $T_{act}$, which is located across the melting region, so to promote only a partial melting; (*iii*) subjecting the specimen to thermal cycles, by cooling below $T_c$ and heating up to $T_{act}$. Step (*i*) and (*ii*) may be considered as a sort of training or programming step, in which the specimen is prepared for the stress-free actuation displayed in step (*iii*). More in detail, in the second protocol, the specimen was heated at 80 °C and subjected to a given level of pre-strain; different levels of the tensile pre-strain were explored: 10%, 20%, 30% and 40%. Then, the specimen was cooled under fixed stress conditions, at the levels of stress that corresponds to the deformation achieved in the previous step, at 2 °C min$^{-1}$ down to −20 °C. The specimen was then unloaded (down to a minimum load equal to 0.001 N) and heated under quasi-stress free conditions at a constant heating rate of 2 °C min$^{-1}$ up to the actuation temperature, $T_{act}$, nominally equal to 43 °C, in order to provide partial melting. Finally, the reversible elongation/contraction effect was measured along a cooling/heating cycle at 2 °C min$^{-1}$ in the -20 °C/80 °C temperature region.

The curves in Fig. 7 represent the different reversible two-way strain recovery response in cycles that mainly differ for the applied strain, *i.e.*, (a,b): 5%, (c,d): 10%, (e,f): 20%, (g): 30%, and (h): 40%. For all these cycles a similar history is displayed: the material, after being deformed above $T_m$ at a given strain level, undergoes a further elongation due to the CIE under the applied stress; at this point the stress is removed and the specimen is heated up to $T_{act}$, undergoing partial melting, and consequently partially recovering the applied strain. Finally, by cooling/heating the specimen between this latter temperature and the lowest one, the sample undergoes an elongation–contraction cycle under no external stress applied.

The difference between the curves recorded for similar applied strain, shown in Figs. 7(a) and 7(b), 7(c) and 7(d), 7(e), and 7(f) is ascribed to the difficulty of an accurate experimental application of the desired value of $T_{act}$. When approaching even slightly different values of $T_{act}$, the degree of partial melting may be different, due to the narrow temperature region of melting. Thus, different thermal contraction is found upon heating (end of step (*ii*)) and, consequently, significantly different elongation/contraction cycles (step (*iii*)). This probably represents the most difficult aspect in operatively controlling the two-way reversible effect of this material.

Thus, although not excluding any of the curves for the experimental-model verification, in order to quantify the effect of the applied pre-strain, the most reasonable comparison should be made for the curves having similar degree of partial melting, which may be hypothesized as the curves in which the stress-free recovery approached about half of the whole recovery process, namely, the curves shown in Fig. 7(a), 7(c), 7(e), 7(g), and 7(h), all representing the results for a different value of applied pre-strain.

It may be observed that, under this condition, significant reversible strain variations may be achieved for all the applied pre-strains, with a difference of about 5%, as shown in Fig. 7(a) for a pre-strain equal to 5%, and of about 10%–12% for the remaining cases, very slightly depending on the applied pre-strain.

Concerning the remaining cases in which, due to a slight variation of $T_{act}$, the partial melting degree is different from half of the process, it may be seen that: for lower values of partial melting, *i.e.*, Figs. 7(b) and 7(d), the reversible strain is still significant, but lower than the case of a larger partial melting degree; by contrast, for larger values of the partial degree of melting (Fig. 7(f)), a larger cycle of reversible strain variation may be obtained.

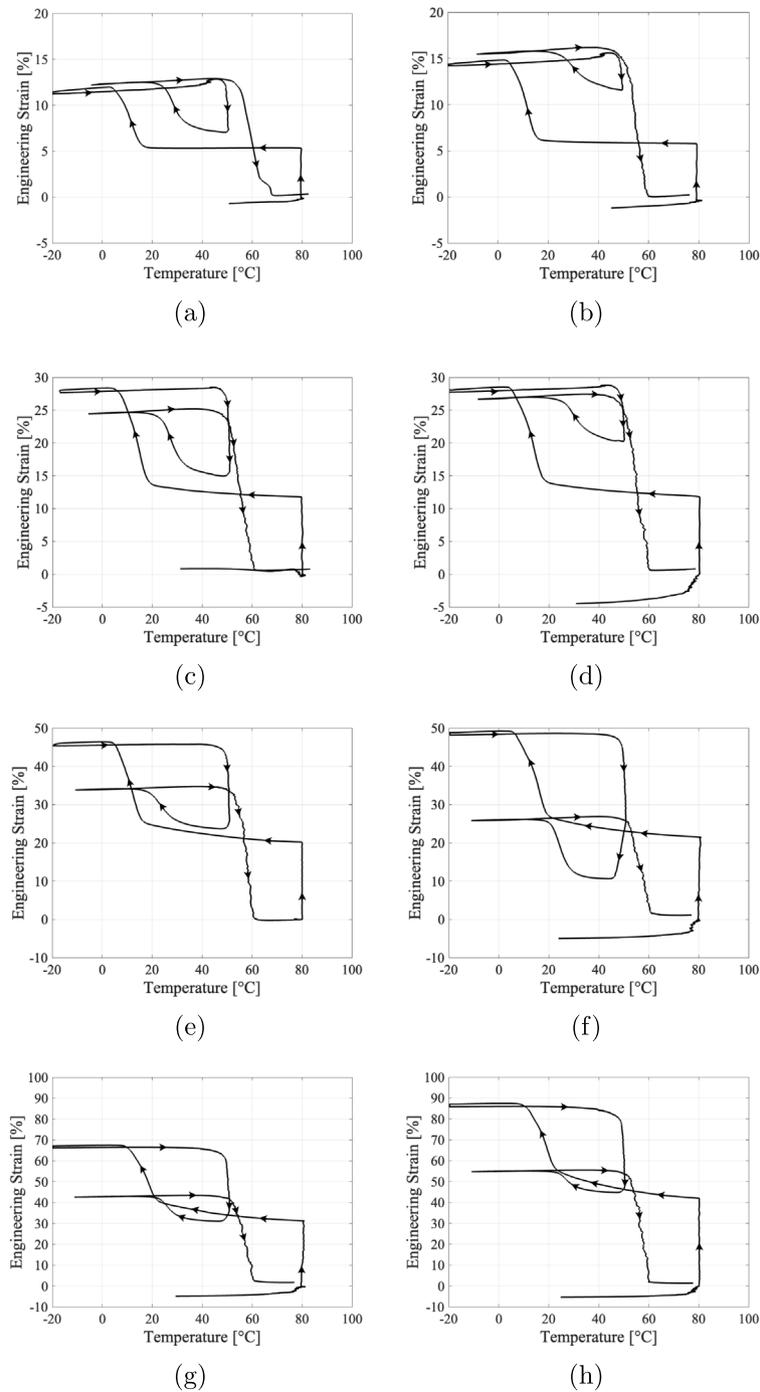

**Fig. 7.** Strain vs temperature curves demonstrating the two-way reversible effect at different levels of applied tensile pre-strain of (a,b): 5%, (c,d): 10%, (e,f): 20%, (g): 30%, and (h): 40%, on the stress-free two-way SME. Arrows indicate the direction in which steps are applied.

The relevant effect of $T_{\text{act}}$ and of the partial degree of melting was further explored in the third protocol test where subsequent stress-free cooling/heating cycles were performed at 2 °C min$^{-1}$ in the $-15°C/T_{\text{act}}$ interval, slightly increasing the value of $T_{\text{act}}$ after each cycle. This approach allowed to effectively collect data concerning the effect of the slight variation of $T_{\text{act}}$ on the two-way reversible response, in spite of the difficulty of controlling precisely this maximum temperature during the cycle. The results are reported for various values of $T_{\text{act}}$ in Fig. 8(a) and 8(b), each graph referring to an individual multi-cycle experiment carried out on two specimens of the same material. Progressively larger partial melting degrees were obtained by slightly increasing $T_{\text{act}}$, and the response of each of these conditions was measured in the corresponding two-way reversible response in heating–cooling cycles.

The results reveal that, for no partial melting there is no reversible effect, as shown by the first highest cooling–heating cycle in Fig. 8(a), which remains almost flat as temperature changes. By increasing $T_{\text{act}}$ a progressive increase in the elongation/contraction effect is shown. This effect may be interpreted as the consequence of having larger and larger portions of amorphous chains that may undergo cooling-induced crystallization under the presence of the internal load. Interestingly, it is also shown that even a very low residual crystallinity may lead to very significant elongations, as shown by the lower curves of Fig. 8(b); for this specimen, the amount of residual crystal may be considered fairly low, but even just this small amount of remaining crystal ensures large reversible contraction-elongation cycles. Furthermore, a shift of the inflection point of the sigmoidal increasing trend of the CIE under stress-free conditions is observed in Fig. 8, due to the presence of the internal load.

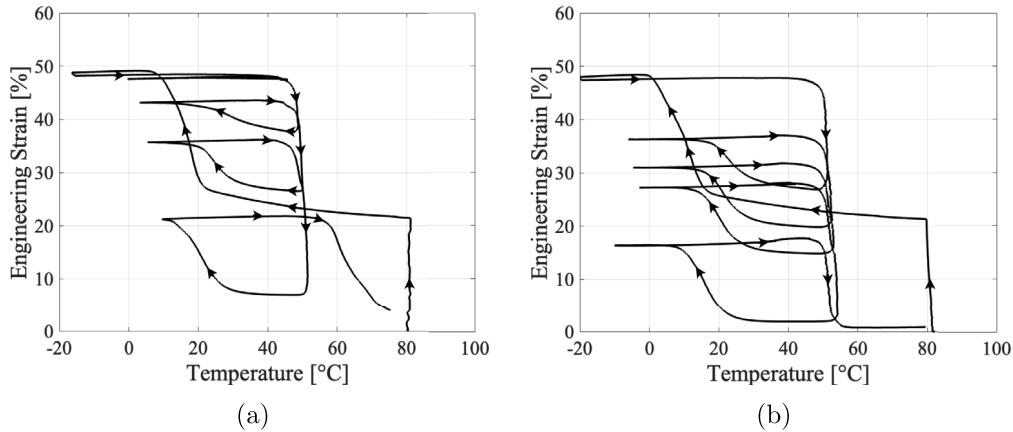

**Fig. 8.** Strain vs temperature curve demonstrating the effect of $T_{act}$ on the stress-free two-way SME carried out on two specimens of the same material. Arrows indicate the direction in which steps are applied.

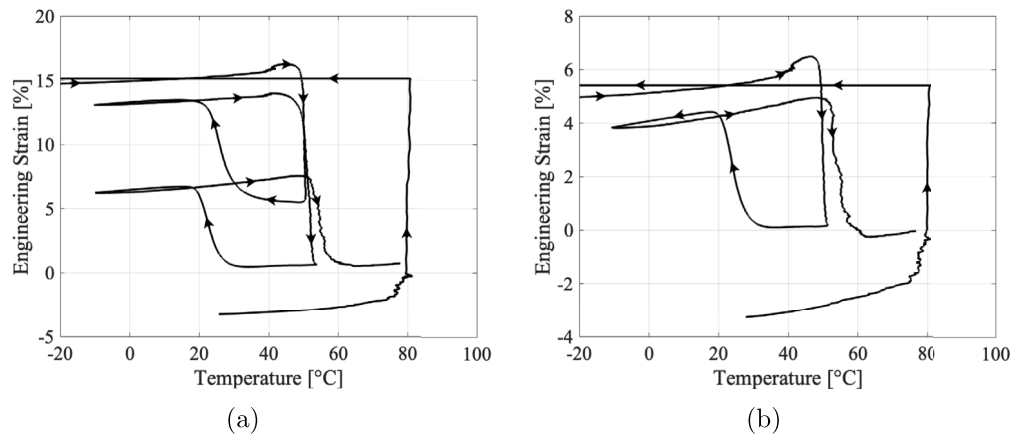

**Fig. 9.** Strain vs temperature curve demonstrating the effect of $T_{act}$ on the stress-free two-way SME. Two pre-strains equal to (a) 15% and (b) 5.5% were considered. Arrows indicate the direction in which steps are applied.

Finally, it is shown that the training/programming history before the cooling–heating cycle may be further simplified, by deforming the specimen and cooling it under fixed strain, in a manner very similar to the first three (four) steps of the one-way cycle (see Section 2.3) that precede the recovery stage, and with no need of a preceding cooling-induced crystallization under fixed stress. Results in Fig. 9 confirm previous observations.

## 3. Physical interpretation of experimental results

In order to highlight the key ingredients necessary for the model formulation, we here provide a physical interpretation of the experimental results presented in Section 2.

The architecture of the investigated semi-crystalline network is schematically depicted in Fig. 10. For temperatures higher than $T_m$, the material consists of the amorphous PCL network (blue random-coiled line), chemically connected by cross-links (orange circles), and behaves as an isotropic rubber. Below $T_c$, the crystalline domains (blue regular segments) start to form and, once crystallization is completed, the material is made of PCL crystallites, connected by the amorphous cross-linked phase, and behaves as an isotropic elasto-plastic solid. When heated again above $T_m$, the material reversibly returns to its amorphous state. The chemical cross-links act as netpoints, connect the network of flexible amorphous PCL chain segments, and determine the permanent shape of the polymer. The crystalline domains act as thermal switches, are generated after cooling the polymer below $T_c$, and are responsible for both the temporary shape fixation and the permanent shape recovery, determined by $T_m$.

Accordingly, a physical interpretation of the experiments performed on the one-way and two-way SMP, under stress and stress-free conditions, can be provided by analyzing material changes at this micro-structural level.

For the one-way SME (Section 2.3), the polymer is heated above $T_m$, yielding high chain mobility. The application of a strain produces macroscopic shape changes (deformed shape) wherein chains align along the loading direction (Wang et al., 2019). The subsequent cooling of the deformed polymer at temperatures lower than $T_c$, and at constant strain, leads to the fixation of the temporary shape and the formation of oriented crystallites. Upon unloading, oriented crystallites impede the recovery of the original shape by stiffening the polymer (shape fixing), allowing to maintain the shape at a progressively lower stress. Heating the deformed polymer above $T_m$ under stress-free conditions, chains attain higher mobility that makes the recovery of the coiled shape imposed by the chemical cross-links possible.

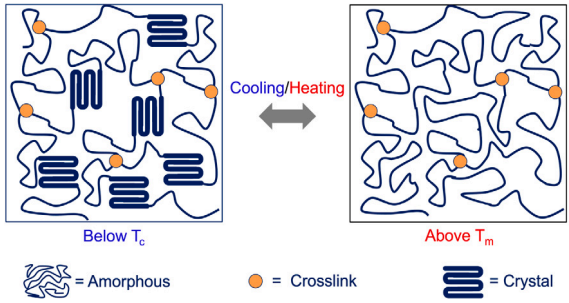

**Fig. 10.** Schematic representation of the architecture of the semi-crystalline network. (For interpretation of the references to color in this figure legend, the reader is referred to the web version of this article.)

For the two-way SME under non-zero stress (Section 2.4.1), the polymer is heated above $T_m$ and subjected to the application of a certain stress, which leads to a pre-stretched (initial) state, prior to cooling–heating cycles. Here, polymer chains are oriented along the loading direction. During cooling, and under fixed stress conditions, for $T_c \leq T < T_m$, in view of the rubber-like elastic behavior, the elastic modulus decreases as temperature decreases, driving the former elongation of the polymer. The subsequent second elongation of the polymer occurs at $T < T_c$, as a consequence of the structural evolution process in which crystal formation, tending to relax the stress, promotes an additional stretching in order to satisfy the constant stress condition: the crystallite formation under stress is possibly accompanied by the orientation of the newly formed crystallites (Chung et al., 2008). Heating above $T_c$ under fixed stress, the configuration displayed before cooling is recovered, as dictated by both the rubber-like elastic behavior and the applied stress. Indeed, the crystalline phase is melted. Moreover, variations of the heating and cooling rates are mainly associated to shifts of the CIE and MIC processes. As discussed in Section 2.4.1, higher heating rates lead to a steeper MIC process and to a shift to higher temperatures of the overall MIC effect; similarly evidenced for the heating step, higher cooling rates lead to a steeper strain increase and to a shift to lower temperatures of the CIE. Rate dependence is thus implicitly included in shifts of the CIE and MIC temperatures.

For the two-way SME under zero stress (Section 2.4.2), similar considerations can be done for the first isothermal loading above $T_m$, and the subsequent cooling below $T_c$. During stress-free heating, macroscopic shape recovery is activated, involving the part of the polymer in which melting occurs. On subsequent cooling, the molten polymer recrystallizes and the chain mobility decreases, so that internal tensile forces become dominant and promote the CIE along the stress direction (Zhou et al., 2014). From this description, and recalling what discussed in Section 1, it is clear that the melted phase acts as actuation phase, whereas the unmelted part is the skeleton phase. The physical background lies in the creation of an internal stress between the partially-melted crystalline phase, which attempts to recover after being generated in the predeformed covalent network, and the remaining undeformed crystalline structure. Indeed, the creation and presence of an internal stress represents the driving force for polymer shape changes during recrystallization, determining an actuation phase that is reversible upon partial melting at heating. The response strongly depends on the relative contents of these two phases, often leading to a non-trivial correspondence between the actuation temperature, $T_{\text{act}}$, and the overall reversible elongation, as demonstrated and discussed in Section 2.4.2. It is frequent that temperature values in the middle of the melting peak are usually required to obtain significant results, since too high or too low values are not able to effectively induce the reversible two-way SME (Yuan et al., 2020).

It is worth highlighting that the unmelted crystals work also as nucleating seeds, thus directing and accelerating the crystallization (Yuan et al., 2020). This effect is referred to as self-nucleated crystallization (SNC) (Michell et al., 2015; Sangroniz et al., 2018), *self-seeds* or *self-nuclei*, and enhances crystallization kinetics, determining a significant increase of the crystallization temperature. A similar effect has been observed in the experimental tests carried out throughout this work, and shown by the curves graphed in Fig. 8. SNC may also involve changes in morphology, such as lamellar thickness (Yuan et al., 2020), in melting temperature, and in crystallization kinetics of a phase in polymorphic materials.

The above discussion highlights that material response depends on variations of the crystalline fraction due to crystallization and melting processes within cooling and heating cycles. Therefore, it is necessary to account for both the morphological micro-structural changes, which occur during each step, and the interaction between the two material phases, into the phenomenological framework for modeling the behavior of semi-crystalline polymer networks. Specifically, it is necessary to provide a comprehensive description of the (*i*) material behavior above $T_m$ and below $T_c$, (*ii*) temperature-dependent evolution of crystallinity during repeated cooling and heating, (*iii*) peculiar cooling-induced elongation (CIE) and melting-induced contraction (MIC) under stress and stress-free conditions, (*iv*) onset of the internal stress, (*v*) effect of additional parameters (*e.g.*, heating/cooling rate, applied load, internal stress, unmelted crystals) on actuation strain and crystallization/melting temperatures.

## 4. Continuum model

This section presents the formulation of the continuum model and its numerical implementation. First, the control variables of the framework are elucidated. Accordingly, phase parameters are introduced to describe the volumetric fractions and the weight contents of the amorphous and crystalline phases, and assumptions on kinematics, in terms of stretch and strain decompositions, are performed. The constitutive equations as well as the evolution laws are then presented. Finally, the numerical treatment of model equations is discussed.

### 4.1. Volumetric fractions

According to the structure of the semi-crystalline network reported in Fig. 10, we introduce two scalar phase parameters defining the volumetric fractions of the amorphous (cross-linked) and the crystalline phases, $\xi_A$ and $\xi_C$, respectively. It holds that

$$\xi_A = \frac{V_A}{V}, \qquad \xi_C = \frac{V_C}{V}, \qquad \xi_A + \xi_C = 1, \qquad (3)$$

with $V = V_A + V_C$ the total volume of the network, and $V_A$ and $V_C$ the volume of the amorphous and crystalline phase, respectively. The volumetric fractions, $\xi_A$ and $\xi_C$, respect the following lower and upper bonds

$$0 < \xi_A \leq 1, \qquad 0 \leq \xi_C \leq \xi_C^{max}, \qquad (4)$$

where $\xi_C^{max} < 1$ is the maximum crystallinity volumetric content. Above the melting temperature, $T_m$, $\xi_A = 1$ and $\xi_C = 0$, meaning that the material is a fully-amorphous chain network; for $T \leq T_c$, the material is a semi-crystalline network, and the crystalline phase achieves its upper bound, therefore implying $\xi_C = \xi_C^{max}$.

In view of the upper bond of the crystalline volumetric fraction, variable $\xi_C$, which is related to the extent of the crystalline transition and the state of the network (Liu et al., 2006), is chosen as *phase variable* to characterize the semi-crystalline network structure. Accordingly, the volumetric fraction of the amorphous phase is conveniently rewritten as $\xi_A = 1 - \xi_C$.

To derive the maximum crystallinity content, $\xi_C^{max}$, we relate the weight content of the crystalline phase, $\xi_C^w$, whose measurement is performed by means of DSC tests (see Fig. 1(b)), to the phase variable $\xi_C$. It holds that

$$\xi_C^w = \frac{m_C}{m_A + m_C} = \frac{\rho_C \xi_C}{\rho_A (1 - \xi_C) + \rho_C \xi_C}, \qquad (5)$$

where $m_A$ and $m_C$ are the masses of the amorphous and crystalline phase, and $\rho_A$ and $\rho_C$ the respective mass densities, which are evaluated as

$$\rho_A = \frac{m_A}{V_A}, \qquad \rho_C = \frac{m_C}{V_C}, \tag{6}$$

yielding the right-hand side term of Eq. (5).

According to Eqs (3)$_2$ and (6), the maximum crystallinity volumetric content can be cast in the form

$$\xi_C^{max} = \frac{V_C^{max}}{V_A^{min} + V_C^{max}} = \frac{\rho_A \xi_C^{w,max}}{\rho_C (1 - \xi_C^{w,max}) + \rho_A \xi_C^{w,max}}. \tag{7}$$

where the maximum crystallinity weight content, $\xi_C^{w,max}$, corresponds to Eq. (5) for $m_C = m_C^{max} = \rho_C V_C^{max}$ and $m_A = m_A^{min} = \rho_A V_A^{min}$.

### 4.2. Kinematics

Let $\varepsilon$ be the total *true* strain in the material, and denote with $\lambda$ the total axial stretch. We base the kinematics of the proposed framework for SMP modeling on the following logarithmic relation

$$\varepsilon = \ln[\lambda], \tag{8}$$

and multiplicative decomposition of the stretch

$$\lambda = \lambda^m \lambda^{th}, \tag{9}$$

with $\lambda^m$ and $\lambda^{th}$ the mechanical and thermal contribution to the total stretch, respectively.

Combination of Eqs (8) and (9) leads to rewrite the total *true* strain, $\varepsilon$, as

$$\varepsilon = \varepsilon^m + \varepsilon^{th}, \tag{10}$$

with $\varepsilon^m$ the (total) mechanical strain and $\varepsilon^{th}$ the thermal strain, whose functional dependence is restricted only to temperature.

Standard arguments in SMP modeling commonly assume the equality between either stresses (Liu et al., 2006; Baghani et al., 2011, 2012; Scalet et al., 2018), or strains (Boatti et al., 2016; Yan et al., 2020), in the amorphous and crystalline phases. We here proceed according to the latter statement, establishing that the mechanical strains in the amorphous and crystalline phases (respectively, $\varepsilon_A^m$ and $\varepsilon_C^m$) are both equal to the (total) mechanical strain, $\varepsilon^m$, i.e.,

$$\varepsilon^m = \varepsilon_A^m = \varepsilon_C^m. \tag{11}$$

In accordance with the schematic representation depicted in Fig. 11, we adopt a simple hyperelastic behavior for the amorphous network, such that the mechanical strain in the amorphous phase is assumed to be purely elastic, and denoted with $\varepsilon_A^{el}$ hereinafter, hence,

$$\varepsilon_A^m = \varepsilon_A^{el}. \tag{12}$$

Rather, elastic and inelastic contributions are accounted for in the crystalline phase, therefore,

$$\varepsilon_C^m = \varepsilon_C^{el} + \varepsilon^{in}, \tag{13}$$

where $\varepsilon_C^{el}$ represents the elastic strain, and the strain inelastic storage, $\varepsilon^{in}$, depends on temperature.

Furthermore, we consider the inelastic contribution, $\varepsilon^{in}$, to the strain in the crystalline phase, $\varepsilon_C^m$, as subjected to the following additive decomposition

$$\varepsilon^{in} = \varepsilon_C^{in} + \varepsilon^{st,f}. \tag{14}$$

Here, $\varepsilon_C^{in}$ accounts for the CIE and the MIC associated to crystallization evolution. Strain $\varepsilon^{st,f}$ represents the amount of deformation induced by the high-temperature loading (see Figs. 5–9) that is stored temporarily (*frozen*) at low temperature during the cooling cycle, before being fully recovered for $T \geq T_m$ after heating (Boatti et al., 2016).

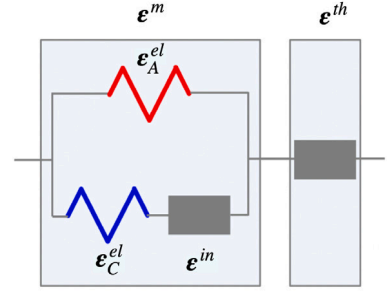

**Fig. 11.** Schematic of the additive decomposition of the total *true* strain.

### 4.3. Constitutive formulation

The constitutive definition of the Helmholtz free energy function is defined by means of a standard rule of mixtures, *viz.*,

$$\psi = (1 - \xi_C) \psi_A + \xi_C \psi_C, \tag{15}$$

with $\psi_A$ and $\psi_C$ the specific free energies (per unit volume) of the amorphous and semi-crystalline network, respectively, and where

$$\psi_\kappa = \psi_\kappa^{el} + \psi_\kappa^{\emptyset}, \tag{16}$$

for $\kappa = A, C$. Terms in Eq. (16) have the following meaning: $\psi_\kappa^{el}$ represents the elastic contribution and $\psi_\kappa^{\emptyset}$ accounts for entropic changes.

The elastic contributions to the amorphous and semi-crystalline network write as

$$\psi_A^{el} = \frac{1}{2} E_A T \varepsilon_A^{el} \varepsilon_A^{el} = \frac{1}{2} E_A T \varepsilon^m \varepsilon^m, \tag{17a}$$

$$\psi_C^{el} = \frac{1}{2} E_C \varepsilon_C^{el} \varepsilon_C^{el} = \frac{1}{2} E_C (\varepsilon^m - \varepsilon^{in})(\varepsilon^m - \varepsilon^{in}), \tag{17b}$$

with $E_A T$ the modulus of the amorphous network, assumed as linearly dependent on the temperature similarly to standard rubbers (Scalet et al., 2018), $E_C$ the modulus of the semi-crystalline network, and where advantage has been taken of Eqs (11), (12) and (13) to replace $\varepsilon_A^{el}$ and $\varepsilon_C^{el}$.

For $\kappa = A, C$, the remaining energy contribution of amorphous and semi-crystalline network is given by

$$\psi_\kappa^{\emptyset} = c_\kappa \left( T - T^{\emptyset} - T \ln \tilde{T} \right) + u_\kappa^{\emptyset} - T \eta_\kappa^{\emptyset}, \tag{18}$$

with $\tilde{T} = T/T^{\emptyset}$, $T^{\emptyset}$ the reference value of the temperature, $c_\kappa$ the heat capacity (per unit volume) of phase $\kappa$, and $u_\kappa^{\emptyset}$ and $\eta_\kappa^{\emptyset}$ the internal energy and entropy (per unit volume), respectively, of phase $\kappa$ at temperature $T^{\emptyset}$.

According to statement (11) and the Voigt model, the Cauchy (*true*) stress is described by the rule of mixture, *i.e.*,

$$\sigma = (1 - \xi_C) \sigma_A + \xi_C \sigma_C, \tag{19}$$

with $\sigma_A$ and $\sigma_C$ stresses in the amorphous and semi-crystalline phases, respectively, and with $\sigma_A \neq \sigma_C$. In view of Eqs (16) and (17), it holds that

$$\sigma_A = \frac{\partial \psi_A^{el}}{\partial \varepsilon_A^{el}} = E_A T \varepsilon_A^{el} = E_A T \varepsilon^m, \tag{20a}$$

$$\sigma_C = \frac{\partial \psi_C^{el}}{\partial \varepsilon_C^{el}} = E_C \varepsilon_C^{el} = E_C (\varepsilon^m - \varepsilon^{in}). \tag{20b}$$

It follows that the Cauchy stress can be rephrased in the form

$$\sigma = E \varepsilon^m - \xi_C E_C \varepsilon^{in} = E (\varepsilon - \varepsilon^{th}) - \xi_C \sigma^{int}, \tag{21}$$

after introduction of the total stiffness modulus,

$$E = (1 - \xi_C) E_A T + \xi_C E_C, \tag{22}$$

and the induced internal stress,

$$\sigma^{int} = E_C \, \varepsilon^{in} = E_C \, (\varepsilon_C^{in} + \varepsilon^{st,f}) \,, \tag{23}$$

to account for the actuation under stress-free conditions.

Moreover, taking advantage of Eq. (21), the total strain, $\varepsilon$, can be expressed in the form

$$\varepsilon = \varepsilon^{th} + \frac{1}{E}(\sigma + \xi_C \, \sigma^{int}) \,, \tag{24}$$

where the thermal strain, $\varepsilon^{th}$, is here taken as

$$\varepsilon^{th} = \ln\left[\lambda^{th}\right] = \ln\left[1 + \left((1 - \xi_C)\,\alpha_A + \xi_C \, \alpha_C\right)\,\Delta T - \xi_C \, \Delta\varepsilon_C^{th}\right] \,, \tag{25}$$

with $\alpha_A$ and $\alpha_C$ thermal expansion coefficients of the amorphous and semi-crystalline network, respectively, $\Delta T = T - T_0$, with $T_0$ the initial temperature, and $\Delta\varepsilon_C^{th}$ the strain increment parameter related to the volumetric contraction (expansion), occurring during crystallization (melting) (Scalet et al., 2018).

To derive the nominal (Piola, Piola–Kirchhoff, or *engineering*) stress, P, consider $A_0$ and $L_0$ as the cross-sectional area and the length of the undeformed specimen, respectively, that are subjected to the uniaxial test, and let $f$ denote the applied axial force through which $A_0$ and $L_0$ are deformed to $A$ and $L$.

The Cauchy (*true*) and the Piola (*engineering*) stresses can be expressed according to the following relations

$$\sigma = \frac{f}{A}, \qquad P = \frac{f}{A_0}. \tag{26}$$

In accordance with experimental evidences, we consider the material as incompressible, so that

$$A_0 \, L_0 \simeq A \, L \,, \tag{27}$$

whence, making use of Eq. (27) to replace $A$ in Eq. (26)$_1$, and taking into account that the stretch writes as $\lambda = L \, L_0^{-1}$, the Piola stress (26)$_2$ turns into

$$P = \sigma \, \lambda^{-1} = \sigma \, \lambda^{-m} \, \lambda^{-th} \,, \tag{28}$$

according to the multiplicative decomposition of the stretch (9).

### 4.4. Evolution laws

According to the transient evolution of the crystallinity content measured in DSC tests (see Figs. 1(b) and 2), we propose the following evolution law to describe the temperature-dependent evolution of the crystalline phase, expressed as volume fraction, $\xi_C$, within a cooling and subsequent heating cycle

$$\dot{\xi}_C = \begin{cases} -\dfrac{\beta^{cool} \, \exp\left[\beta^{cool}\,(T - T_{c,eff})\right]}{\left(1 + \exp\left[\beta^{cool}\,(T - T_{c,eff})\right]\right)^2} \, \xi_C^0 \, \dot{T} \,, & \text{if } \dot{T} < 0 \,, \\[2ex] -\dfrac{\beta^{heat} \, \exp\left[\beta^{heat}\,(T - T_{m,eff})\right]}{\left(1 + \exp\left[\beta^{heat}\,(T - T_{m,eff})\right]\right)^2} \, \xi_C^0 \, \dot{T} \,, & \text{if } \dot{T} > 0 \,, \\[2ex] 0, & \text{if } \dot{T} = 0 \,. \end{cases} \tag{29}$$

Such evolution law allows the description of the cyclic transformation of the crystalline volume fraction within multiple heating and cooling steps.

Symbols in system (29) have the following meaning. Parameters $\beta^{cool}$ and $\beta^{heat}$ are positive material constants for the cooling and heating cycle, respectively. The volume fraction, $\xi_C^0$, is given by the following definition

$$\xi_C^0 = \begin{cases} \xi_C^{max} - \xi_C^{heat} \,, & \text{if } \dot{T} \leq 0 \,, \\ \xi_C^{cool} \,, & \text{if } \dot{T} > 0 \,, \end{cases} \tag{30}$$

with $\xi_C^{heat}$ and $\xi_C^{cool}$ representing the fraction of crystalline phase available from previous heating and cooling cycle, respectively, and where, for initialization purposes,

$$\xi_C(t=0) = \frac{\xi_C^{max}}{1 + \exp\left[\beta^{cool}\,(T - T_c)\right]} \,. \tag{31}$$

It is noted that parameter $\xi_C^0$ is introduced to ensure the cyclic description and the continuity of the law. In fact, it is equal to the values of the crystalline volume fraction at reversal points, *i.e.*, when a passage from heating to cooling (or viceversa) takes place.

As discussed in Section 2, temperatures $T_{c,eff}$ and $T_{m,eff}$ are the effective crystallization and melting temperatures, introduced for taking into account that the crystallization process may be (*i*) affected by the application of an external stress in the two-way shape memory tests, (*ii*) conditioned by the internal stress and the melted/unmelted phase during stress-free shape memory tests, and (*iii*) kinetically delayed by an increase of the cooling rate. Rather, the melting process may be kinetically delayed by an increase of the heating rate. They correspond to the inflection point of the sigmoidal curves and are defined empirically as

$$T_{c,eff} = T_c + \tau_c \, \nu^{cool} + \beta_C \, |\sigma_{end}| + \delta_C \, |\sigma_{end}^{int}| + \gamma_C (1 - \xi_C^{heat}) \tag{32a}$$

$$T_{m,eff} = T_m + \tau_m \, \nu^{heat}. \tag{32b}$$

Here, $\tau_c < 0$ and $\tau_m > 0$ describe the linear shift in the transition temperatures (Scalet et al., 2018) depending on the cooling and heating rate, $\nu^{cool}$ and $\nu^{heat}$, respectively. Parameter $\beta_C > 0$ accounts for the stress influence on the crystallization process, being $\sigma_{end}$ the value of the stress at the end of the isothermal loading, whereas $\delta_C > 0$ describes the shift of transformation temperature due to the internal stress (23), being $\sigma_{end}^{int}$ the internal stress generated at the end of the previous heating cycle. The introduction of $\gamma_C$ allows the description of the dependence of $T_{c,eff}$ on the crystalline phase.

Lastly, in view of Eqs (11), (13), and (14), we provide the evolution equations for the strain induced by the formation of crystals, $\varepsilon_C^{in}$, and for the deformation induced by the high-temperature loading, which is stored temporarily at low temperature, $\varepsilon^{st,f}$.

The evolution for $\varepsilon_C^{in}$ is here chosen in the form (Scalet et al., 2018)

$$\dot{\varepsilon}_C^{in} = \begin{cases} \dot{\xi}_C \, \dfrac{\sigma}{\alpha} + \dot{\xi}_C \, \gamma \, \sigma_{end}^{int} + \dot{\xi}_C \, \theta \, \dfrac{\sigma_{end}^{int}}{|\sigma_{end}^{int}|} \,, & \text{if } \dot{T} < 0 \,, \\[2ex] \dfrac{\dot{\xi}_C}{\xi_C} \, \varepsilon_C^{in}, & \text{if } \dot{T} > 0 \,, \\[2ex] 0, & \text{if } \dot{T} = 0 \,, \end{cases} \tag{33}$$

where the term $\sigma/\alpha$ accounts for the contribution of the applied stress, $\gamma \, \sigma_{end}^{int}$ for the contribution of the internal stress generated at the end of the previous heating cycle, and where

$$\theta = \theta_C \, \xi_C^{heat} + \theta_A (1 - \xi_C^{heat}) \tag{34}$$

considers the effect of the amorphous and crystalline phase, by means of $\theta_A$ and $\theta_C$, respectively, at the end of the previous heating cycle.

The strain $\varepsilon^{st,f}$ stores the deformation induced by the applied load, varying under isothermal loading, for $T > T_m$. During heating under stress-free conditions, $\varepsilon^{st,f}$ is expected to decrease, while it is kept constant during heating under an applied stress. Accordingly, the evolution law aiming at describing the *frozen* strain is here chosen as

$$\dot{\varepsilon}^{st,f} = \begin{cases} \beta \, \dot{\varepsilon}_A^{el} \,, & \text{if } \dot{T} = 0 \text{ and } T > T_m \,, \\[1ex] \dfrac{\dot{\xi}_C}{\xi_C} \, \varepsilon^{st,f} \,, & \text{if } \dot{T} > 0 \text{ and } |\sigma| = 0 \,, \\[1ex] 0, & \text{otherwise} \,, \end{cases} \tag{35}$$

with $\beta$ a positive material parameter.

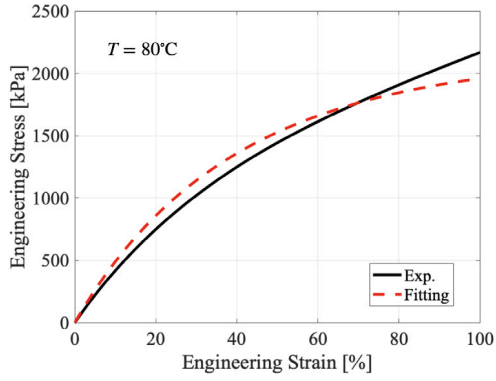

**Fig. 12.** Identification of parameter $E_A$ on stress–strain curve for the amorphous material.

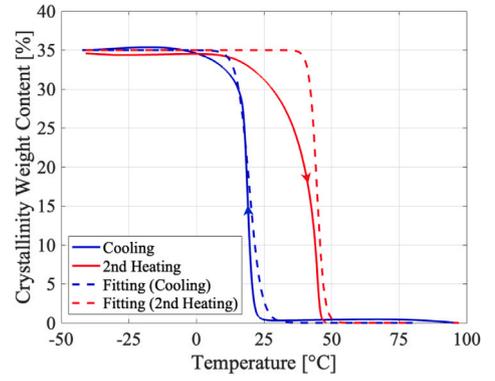

**Fig. 13.** Identification of parameters $\xi_C^{w,max}$, $\xi_C^{max}$, $\beta^{cool}$, and $\beta^{heat}$ on crystalline content-temperature curves.

## 4.5. Model implementation

We now focus on the numerical treatment of model equations. For the sake of notation simplicity, we use subscripts $(n)$ and $(n+1)$ for all the quantities evaluated at previous time $t_{(n)}$ and current time $t_{(n+1)}$, respectively.

We employ a backward-Euler integration algorithm for the evolution Eq. (33) of the strain $\varepsilon_C^{in}$ induced by crystallization

$$\varepsilon_{C(n+1)}^{in} = \begin{cases} \varepsilon_{C(n)}^{in} + \left[\xi_{C(n+1)} - \xi_{C(n)}\right] \left[\dfrac{\sigma_{(n+1)}}{\alpha} + \gamma\, \sigma_{end}^{int} + \theta\, \dfrac{\sigma_{end}^{int}}{|\sigma_{end}^{int}|}\right], & \text{if } T_{(n+1)} - T_{(n)} < 0, \\ \dfrac{\xi_{C(n+1)}}{\xi_{C(n)}} \varepsilon_{C(n)}^{in}, & \text{if } T_{(n+1)} - T_{(n)} > 0, \\ \varepsilon_{C(n)}^{in}, & \text{if } T_{(n+1)} - T_{(n)} = 0, \end{cases}$$

(36)

as well as for the evolution Eq. (35) of the frozen strain $\varepsilon^{st,f}$

$$\varepsilon_{(n+1)}^{st,f} = \begin{cases} \varepsilon_{(n)}^{st,f} + \beta \left[\varepsilon_{A(n+1)}^{el} - \varepsilon_{A(n)}^{el}\right], & \text{if } T_{(n+1)} - T_{(n)} = 0 \text{ and } T_{(n+1)} > T_m, \\ \dfrac{\xi_{C(n+1)}}{\xi_{C(n)}} \varepsilon_{(n)}^{st,f}, & \text{if } T_{(n+1)} - T_{(n)} > 0 \text{ and } |\sigma_{(n+1)}| = 0, \\ \varepsilon_{(n)}^{st,f}, & \text{otherwise}. \end{cases}$$

(37)

Finally, the evolution equation for the crystalline phase (29) is discretized as follows

$$\xi_{C(n+1)} = \xi_{C(n)} + \begin{cases} -\dfrac{\beta^{cool}\, \exp\left[\beta^{cool}(T_{(n+1)} - T_{c,eff})\right]}{\left(1 + \exp\left[\beta^{cool}(T_{(n+1)} - T_{c,eff})\right]\right)^2}\, \xi_C^0\, \left[T_{(n+1)} - T_{(n)}\right], \\ \qquad\qquad\qquad\qquad\qquad\qquad\qquad \text{if } T_{(n+1)} - T_{(n)} \le 0, \\[4pt] -\dfrac{\beta^{heat}\, \exp\left[\beta^{heat}(T_{(n+1)} - T_{m,eff})\right]}{\left(1 + \exp\left[\beta^{heat}(T_{(n+1)} - T_{m,eff})\right]\right)^2}\, \xi_C^0\, \left[T_{(n+1)} - T_{(n)}\right], \\ \qquad\qquad\qquad\qquad\qquad\qquad\qquad \text{if } T_{(n+1)} - T_{(n)} > 0. \end{cases}$$

(38)

All the remaining model equations are evaluated at current time $t_{(n+1)}$.

The solution to the non-linear system is obtained by means of the function *fsolve* implemented in the optimization toolbox of the program Matlab. It is worth highlighting that the model can be easily implemented and allows for a direct and quick prediction of the shape-memory behavior in semi-crystalline networks.

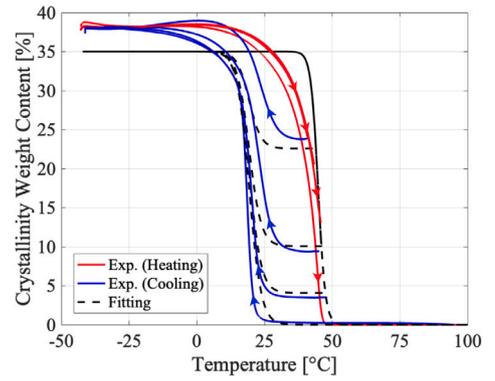

**Fig. 14.** Model validation on cyclic crystalline-temperature curves.

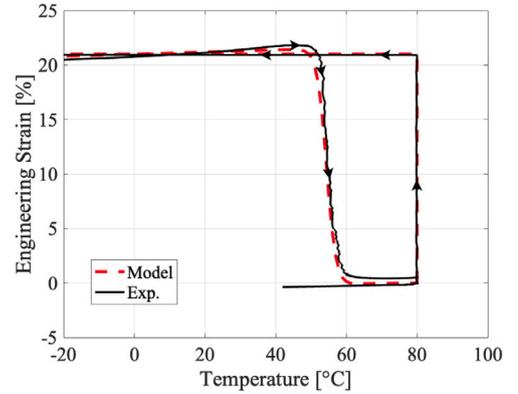

**Fig. 15.** Model validation on experimental *engineering* strain vs temperature curves demonstrating the one-way SME.

## 5. Numerical versus experimental results

We now present the results of the performed numerical simulations and a comparison with the experimental campaign in order to validate the proposed one-dimensional phenomenological continuum framework.

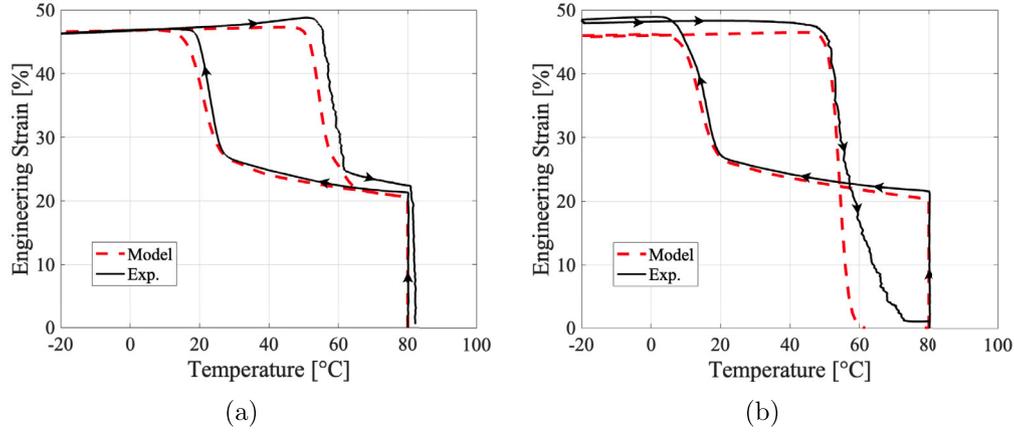

**Fig. 16.** Model validation on experimental *engineering* strain vs temperature curves, demonstrating the two-way SME under (a) constant applied stress and (b) stress-free recovery.

*5.1. Model calibration*

The proposed model presents the following material parameters: (*i*) two parameters for the stiffness moduli of the amorphous and crystalline phases, $E_A$ and $E_C$; (*ii*) the crystallization and melting temperatures, $T_c$ and $T_m$; (*iii*) the crystalline phase evolution parameters, $\rho_A$, $\rho_C$, $\xi_C^{w,max}$, $\xi_C^{max}$, $\beta^{cool}$, and $\beta^{heat}$; (*iv*) the thermal expansion coefficients, $\alpha_A$ and $\alpha_C$, and the thermal strain increment, $\Delta\varepsilon_C^{th}$; (*v*) the heating and cooling rate parameters, $\tau_m$ and $\tau_c$; (*vi*) the evolution parameters, $\alpha$, $\theta_A$, $\theta_C$, $\gamma$ and $\beta$; (*vi*) temperature parameters, $\beta_C$, $\delta_C$ and $\gamma_C$. All the listed parameters have a physical interpretation or can be intuitively derived via standard tests.

The mechanical parameters related to the hyperelastic responses of the amorphous and crystalline phases, *i.e.*, $E_A$ and $E_C$, can be derived from tensile tests above the melting and below the crystallization temperature, $T_m$ and $T_c$, respectively. Fig. 12 shows the identification process of $E_A$ on the experimental curves described in Section 2.2. In the present case, $E_C$ is identified to catch the elastic recoil in the one-way shape memory curve (see Fig. 5).

The crystallization and melting temperatures, $T_c$ and $T_m$, can be calibrated from a DSC scan, as in the present case (see Section 2.2), or, alternatively, from crystallinity content versus temperature curves.

Crystalline phase evolution parameters, $\xi_C^{w,max}$, $\xi_C^{max}$, $\beta^{cool}$, and $\beta^{heat}$, can be calibrated from weight crystalline content versus temperature curves measured in DSC analysis (see Section 2.2). To complete the identification, densities $\rho_A$ and $\rho_C$ are taken from Ketelaars et al. (1997). Fig. 13 presents the results of the identification process on experimental curves described in Fig. 1(b), after the application of Eq. (5).

To verify the evolution law for the crystalline phase adopted in the present framework, numerical and experimental (see Fig. 2(b)) curves are compared in Fig. 14, showing very good agreement and confirming the suitability of the adopted law.

The thermal expansion coefficients, $\alpha_A$ and $\alpha_C$, and the thermal strain increment, $\Delta\varepsilon_C^{th}$, due to crystallization, can be calibrated from a cooling/heating curve at zero stress or, alternatively, from a two-way shape-memory strain-temperature curve at low applied stress, where the thermal expansion and contraction effects dominate material response. In the absence of these data, they can be deduced from the two-way curve in Fig. 6(a), since the slight strain change before melting is associated to thermal expansion, as discussed in Sections 2.3 and 2.4.

Heating and cooling rate parameters, $\tau_m$ and $\tau_c$, can be calibrated from shape-memory tests performed at different cooling and heating rates, respectively. For the case at hand, they are calibrated on the two-way curve in Fig. 6(a) and validated against the other curves where different rates were employed.

**Table 1**
Model parameters adopted in the numerical simulations.

| Parameter | Value | Units |
|---|---|---|
| $E_A$ | 0.016 | MPa/K |
| $E_C$ | 2895 | MPa |
| $T_c$ | 19 | °C |
| $T_m$ | 45 | °C |
| $\rho_A$ | 1.081 | g/cm$^3$ |
| $\rho_C$ | 1.195 | g/cm$^3$ |
| $\xi_C^{w,max}$ | 0.35 | - |
| $\xi_C^{max}$ | 0.3275 | - |
| $\beta^{cool}$ | 0.45 | °C$^{-1}$ |
| $\beta^{heat}$ | 0.75 | °C$^{-1}$ |
| $\alpha_C$ | $1.5 \cdot 10^{-4}$ | °C$^{-1}$ |
| $\alpha_A$ | $0.5 \cdot 10^{-4}$ | °C$^{-1}$ |
| $\Delta\varepsilon_C^{th}$ | $0.8 \cdot 10^{-2}$ | - |
| $\tau_m$ | 5 | min |
| $\tau_c$ | −2.3 | min |
| $\alpha$ | 4.2 | MPa |
| $\theta_A$ | 0.15 | - |
| $\theta_C$ | 1.6 | - |
| $\beta$ | 1.25 | - |
| $\gamma$ | 0.00035 | MPa$^{-1}$ |
| $\beta_C$ | 6 | °C/MPa |
| $\delta_C$ | 0.02 | °C/MPa |
| $\gamma_C$ | 5.5 | °C |

Evolution parameters $\alpha$ and $\beta$ have no direct physical meaning. However, since they are associated to the evolution of the inelastic strains during non-zero stress cooling and isothermal loading, they are calibrated on the two-way curve reported in Fig. 6(a). Parameters $\gamma$, $\theta_A$ and $\theta_C$, associated to stress-free cooling, are deduced from the corresponding stress-free curves in Fig. 7 (e,f).

Similarly, effective crystallization temperature parameters $\beta_C$, $\delta_C$ and $\gamma_C$ can be derived from the corresponding stress-free curves in Fig. 7 (e,f).

Table 1 contains all the identified parameters of the model.

*5.2. One-way shape-memory tests*

According to the experimental procedure described in Section 2.3, we here show the simulation of the one-way shape-memory tests. We replicate such procedure by keeping the specimen above $T_m$, providing a deformation up to a certain *engineering* strain, $\varepsilon_{app} = \lambda - 1$, and cooling the deformed specimen below $T_c$ at 2 °C/min, while $\varepsilon_{app}$ is kept constant. The recovery behavior is simulated by heating the deformed specimen above $T_m$ at 2 °C/min under stress-free conditions.

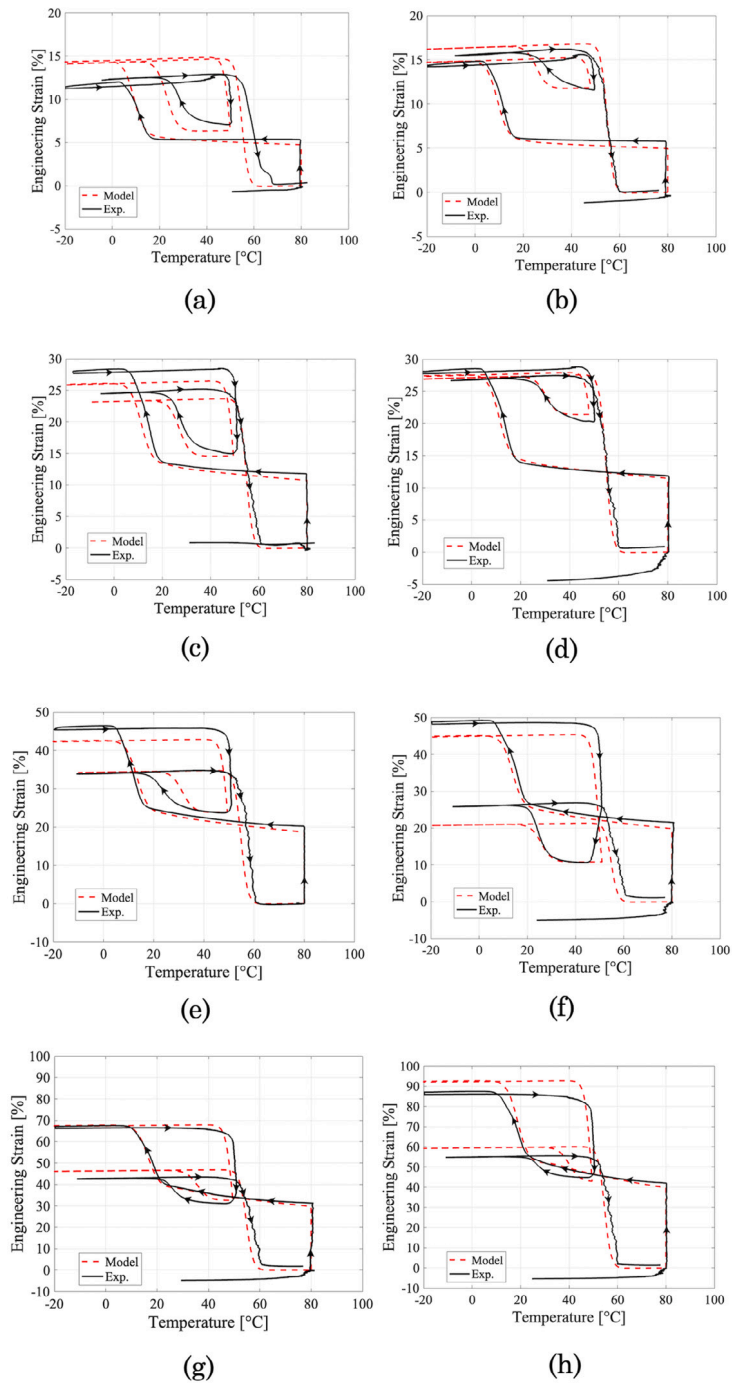

**Fig. 17.** *Engineering* strain vs temperature experimental (black-solid) and model (red-dashed) curves demonstrating the two-way reversible effect at different levels of applied tensile pre-strain, (a,b): 5%, (c,d): 10%, (e,f): 20%, (g): 30%, and (h): 40%, on the stress-free two-way SME. (For interpretation of the references to color in this figure legend, the reader is referred to the web version of this article.)

Fig. 15 shows the strain evolution as a function of temperature under the consideration of an applied strain of 21%, comparing model results with the experimental result previously shown in Fig. 5.

The results demonstrate very good agreement between experiments and model predictions in terms of shape fixity, shape recovery, and the effect of thermal expansion on the slight strain change before melting.

*5.3. Two-way shape-memory behavior under applied stress*

We then simulate the two-way shape-memory tests, according to the experimental procedure described in Section 2.4.1 and the results shown in Fig. 6. The procedure is replicated by keeping the specimen above the melting temperature, $T_m$, and providing a deformation by means of the application of different values of *engineering* stress. A cooling–heating cycle is then performed on the deformed specimen at 2 °C/min, while the applied stress is kept constant.

Fig. 16(a) shows the evolution of the *engineering* strain as a function of temperature. A stress corresponding to the 20% of strain is applied, and the obtained result is compared with the corresponding experimental result shown in Fig. 6(a).

To complete the discussion, Fig. 16(b), compared with Fig. 6(b), shows the *engineering* strain evolution as a function of temperature, first cooled under an applied stress corresponding to the 20% of strain at 5 °C/min, and then heated at 2 °C/min under stress-free conditions. As it can be observed, also for these tests, there is a good agreement between the curves.

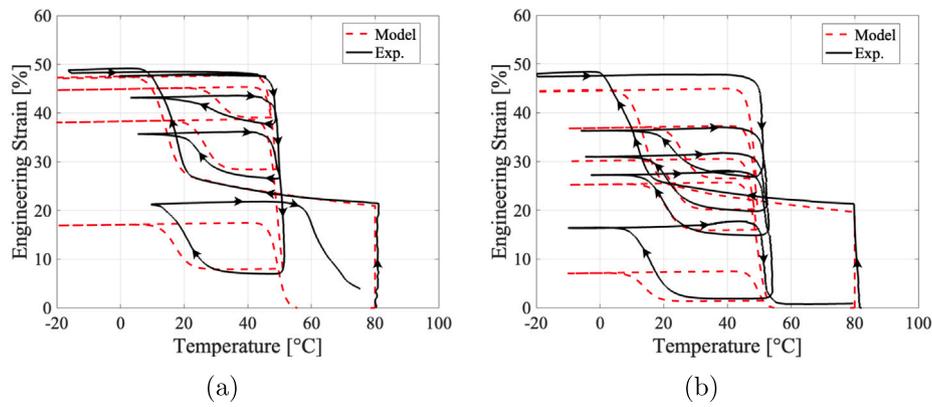

**Fig. 18.** *Engineering* strain vs temperature experimental (black-solid) and model (red-dashed) curves demonstrating the effect of $T_{act}$ on the stress-free two-way SME. (For interpretation of the references to color in this figure legend, the reader is referred to the web version of this article.)

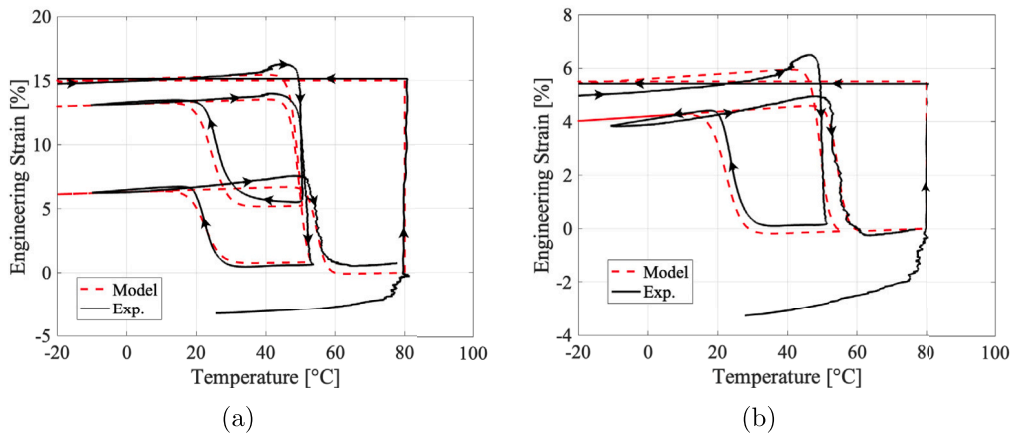

**Fig. 19.** *Engineering* strain vs temperature experimental (black-solid) and model (red-dashed) curves demonstrating the effect of $T_{act}$ on the stress-free two-way SME for an applied tensile pre-strain of (a) 15%, (b) 5.5%. (For interpretation of the references to color in this figure legend, the reader is referred to the web version of this article.)

*5.4. Two-way shape-memory behavior under stress-free conditions*

We present the simulations performed on the two-way shape-memory tests, according to the experimental procedure described in Section 2.4.2 and the results shown in Figs. 7 and 8. The procedure is reproduced by keeping the specimen above $T_m$, deforming via the application of a certain *engineering* stress (corresponding to a certain tensile pre-strain), and by applying on the deformed specimen a cooling step up to $T \ll T_c$ at 5 °C/min, while maintaining the applied stress constant, followed by a heating step up to $T_{act}$ at 1 °C/min, and subsequent cooling–heating cycles under stress-free conditions at a cooling rate of 2 °C/min and heating rate of 2 °C/min.

Figs. 17 and 18 report the *engineering* strain evolution as a function of temperature, in comparison with the experimental results shown in Figs. 7, 8, and 9, respectively. The model predicts well all the phenomena characterizing the two-way SME under stress-free conditions, in particular the actuation strain magnitude and crystallization process onset and termination, especially considering the complex material behavior, material response variability, and the difficulties in the experimental setup and testing. Main differences in some of the subfigures presented in Figs. 17 and 18 can be explained by noting all different cases we are predicting with the same model and the same parameters. That is, by means of the same model, after having properly calibrated the model parameters, we are reproducing a wide range of experimental tests. Moreover, parameters have been fitted with a trial-an-error procedure, but could be optimized with an optimization procedure that is out of the present paper scope. In addition, differences recorded for the applied strain of 5% (Figs. 17 (a-b)) can be ascribed to the experimental difficulties already detailed in Section 2.4.2.

To complete the discussion, the experimental tests, whose results are reported in Fig. 9, are simulated. The procedure is reproduced by keeping the specimen above $T_m$, deforming via the application of a certain *engineering* strain, and by applying on the deformed specimen a cooling step up to $T \ll T_c$ at 5 °C/min, while maintaining the applied strain constant, followed by a heating step up to $T_{act}$ at 1 °C/min, and subsequent cooling–heating cycles under stress-free conditions at a cooling rate of 2 °C/min and heating rate of 2 °C/min. Results are reported in 19, showing again a very good match between model and experimental results.

**6. Conclusions**

In this work, a new one-dimensional macroscopic model based on a phase transition approach has been developed to describe the shape memory behavior of semi-crystalline networks under different thermo-mechanical conditions.

A broad experimental campaign on a semi-crystalline PCL-based network was first performed, allowing to characterize its mechanical and thermal properties, one-way shape-memory behavior, and two-way shape memory behavior under both stress-free and non-zero stress conditions.

A physical interpretation of the experimental results was then provided in order to highlight the key ingredients to formulate the proposed one-dimensional continuum phenomenological model.

The phenomenological framework introduced in this work has been revealed to be of easily implementation and to allow a quick and accurate prediction of the shape-memory behavior in semi-crystalline networks. In fact, numerical results have demonstrated that the model is able to quantitatively describe several important aspects of material shape-memory behavior observed experimentally.

In particular, the conducted experimental and numerical investigation allowed to analyze and discuss all the phenomena and thermomechanical parameters that impact the stress-free two-way SME, in addition to the better known one-way and non-zero stress two-way SME. Accordingly, the crystallinity evolution in cyclic heating–cooling tests was analyzed and shown to determine a strong influence on the mechanical and shape memory behavior. In fact, for the two-way SME under zero stress conditions, results revealed no reversible effects for no partial melting, and a progressive increase in the elongation/contraction effect by increasing $T_{act}$, due to the presence of a larger portion of amorphous chains that may undergo CIE under the presence of the internal load. The amount of crystals, together with the generated internal stress and cooling rate, has been shown to determine also a shift of the crystallization temperature. Significant reversible strain variations were also achieved for all the applied pre-strains, first increasing with increasing pre-strains and then stabilizing.

Additionally, numerical results have allowed to define the ranges of reliability of the model for the material system used for validation; particularly, the model provides results in agreement with experimental evidence for strain ranges up to around 80% and temperature ranges between −20 and 80 °C. Moreover, it is worth highlighting that the model is general and can be applied to any single-component semi-crystalline polymer network exhibiting the one-way SME as well as the two-way SME under stress and stress-free conditions. The choice of focusing on these PCL-based systems is due to the fact that we were able to provide a comprehensive set of experimental data necessary to formulate and validate the model. A further work will be devoted to extend the current proposed one-dimensional model in a three-dimensional setting.

## CRediT authorship contribution statement

**Matteo Arricca:** Methodology, Software, Writing – original draft, Writing – review & editing. **Nicoletta Inverardi:** Conceptualization, Investigation, Methodology, Writing – review & editing. **Stefano Pandini:** Conceptualization, Investigation, Methodology, Writing – original draft, Writing – review & editing. **Maurizio Toselli:** Conceptualization, Investigation, Methodology, Writing – original draft. **Massimo Messori:** Conceptualization, Writing – review & editing. **Ferdinando Auricchio:** Conceptualization, Methodology, Writing – review & editing. **Giulia Scalet:** Conceptualization, Data curation, Funding acquisition, Methodology, Software, Supervision, Writing – original draft, Writing – review & editing.

## Declaration of competing interest

The authors declare that they have no known competing financial interests or personal relationships that could have appeared to influence the work reported in this paper.

## Data availability

## Acknowledgments

This work was funded by the European Union ERC CoDe4Bio Grant ID 101039467. Views and opinions expressed are however those of the author(s) only and do not necessarily reflect those of the European Union or the European Research Council. Neither the European Union nor the granting authority can be held responsible for them.